\documentstyle[12pt]{article}

\begin{document}

\title{The Meaning of Elements of Reality and Quantum Counterfactuals
  -- Reply to Kastner.}

\author{ Lev Vaidman}
\date{}
\maketitle

\begin{center}
{\small \em School of Physics and Astronomy \\
Raymond and Beverly Sackler Faculty of Exact Sciences \\
Tel Aviv University, Tel-Aviv 69978, Israel. \\}
\end{center}

\vspace{2cm}
\begin{abstract}
This paper is the answer to  the
 paper by Kastner [ {\it Found. Phys.} to be published,  quant-ph/9807037]
in which she continued the criticism of the counterfactual usage of the
Aharonov-Bergman-Lebowitz rule in the framework of the time-symmetrized quantum theory,  in
particular, by analyzing the three-box ``paradox''.
It is argued that the criticism is not sound. 
Paradoxical features of the three-box example has been explained.
It is  explained that the elements of reality in the framework of
time-symmetrized quantum theory are counterfactual statements and,
therefore, even conflicting elements of reality can be associated with a
single particle. It is  shown how such ``counterfactual''
elements of reality can be useful in the analysis of a physical
experiment (the three-box example). The validity of Kastner's
application of the consistent histories approach to the
time-symmetrized counterfactuals is questioned. 
\end{abstract}


\section{Elements  of Reality}

Quantum theory teaches us that the concepts of ``reality'' developed
on the basis of the classical physics are not adequate for describing
our world.  A new language with concepts which {\it are} appropriate
is not developed yet and this is probably the root of numerous
controversies regarding interpretation of quantum formalism.  It seems
to me that philosophers of science can make a real contribution for
progress of quantum theory through developing of an appropriate
language. A necessary condition for a success of this wisdom is that
physicists and philosophers will try to understand each other.  I
hope, that the resolution of the current controversy about the
time-symmetrized quantum theory (TSQT) will contribute to such
understanding.

I took part in the development of the TSQT \cite{ABL,AV90,AV91} and I
believe that this is an important and useful formalism. It already
helped us to find several peculiar quantum phenomena tested in
laboratories in the world \cite{w-exp1,w-exp2}. In the framework of
the TSQT I have used terms such as ``elements of reality''
\cite{Va-er,Va-wer} in a sense which seems to be radically different
from the concept of reality considered by philosophers and, apparently,
this is the main reason for  the current
controversy.

I {\it define} that there is an element of reality at time $t$ for an
observable $C$, ``$C=c$'' when it can be inferred with certainty that
the result of a measurement of $C$, if performed, is $c$. Frequently,
in such a situation it is said that the observable $C$ has the value
$c$. It is important to stress that both expressions do not assume
``ontological'' meaning for $c$, the meaning according to which the
system has some (hidden) variable with the value $c$. I do not try to
restore realistic picture of classical theory: in quantum theory
observables do not possess values.  The only meaning of the
expressions: ``the element of reality $C=c$'' and ``$C$ has the value
$c$'' is the operational meaning: it is known with certainty that if
$C$ is measured at time $t$, then the result is $c$.

Clearly, my concept of elements of reality  has its
roots in ``elements of reality'' from the Einstein, Podolsky, and Rosen
paper (EPR) \cite{EPR}. There are numerous works analyzing the EPR elements of
reality. My impression that EPR were looking for an ontological
concept and their ``criteria for elements of reality'' is just a
property of this concept. I had no intention to define such ontological
concept. I apologize for taking this name and using it in a very
different sense, thus, apparently, misleading many readers. I hope to
clarify my intentions here and I welcome suggestions for alternative
name for my concept which will avoid the confusion.

I consider elements of reality as counterfactual statements. Even if  at
time $t$ the system undergoes an interaction with a  measuring device
which measures  $C$, the truth of ``$C=c$'' is ensured not by the final
reading of the pointer of this measurement, but by a counterfactual
statement that if another measurement, with as short duration as we want, is
performed at time $t$, it invariably reads $C=c$.

\section{The three-box example}

The {\em actual} story:

(i) A macroscopic number $N$ of particles (gas) were all prepared at
$t_1$ in a superposition of being in three separated boxes:
\begin{equation}
|\psi_1\rangle = {{1\over \sqrt
  3}}(|A\rangle + |B\rangle + |C\rangle) ,
\end{equation}
with obvious notation: $|A\rangle$ is the state of a particle in box
$A$, etc.

(ii) At later time $t_2$ all the particles were found in another
superposition (this is extremely rare event):
\begin{equation}
|\psi_2\rangle = {{1\over \sqrt
  3}}(|A\rangle + |B\rangle - |C\rangle).
\end{equation}

(iii) In between, at time $t$, {\it weak measurements} of a number of
particles in each box, which are, essentially,  usual measurements of
pressure in each box, have been performed. The readings of the measuring devices
for the pressure in the boxes $A$, $B$ and $C$ were
\begin{eqnarray}
\label{p-p}
\nonumber
p_A = p,\\
p_B = p,\\
\nonumber 
p_C = -p,
\end{eqnarray}
where $p$ is the pressure which is expected to be in a box with $N$
particles.

I am pretty certain that this ``actual'' story never took place
because the probability for successful post-selection (ii) is of the
order of $3^{-N}$; for a macroscopic number $N$ it is too small for
any real chance to see it happens. However, given that the
post-selection (ii) does happen, I am safe to claim that (iii) is
correct, i.e., the measurements of pressure at the intermediate time
with very high probability  yielded the results (\ref{p-p}).

The description of this example in the framework of the time
symmetrized quantum formalism is as follows. Each particle at time $t$
is described by the two-state vector
\begin{equation}
\langle \psi_2|~|\psi_1\rangle = {{1\over 
  3}}
(\langle A| + \langle B|- \langle C|)~ (|A\rangle + |B\rangle + |C\rangle) ,
\end{equation}
The system of all particles (signified by index $i$) is 
 described by the two-state vector
\begin{equation}
\langle \Psi_2|~|\Psi_1\rangle = {{1\over 
  3^N}}\prod_{i=1}^{i=N}
(\langle A|_i + \langle B|_i - \langle C|_i)~ \prod_{i=1}^{i=N}(|A\rangle_i + |B\rangle_i + |C\rangle_i) 
\end{equation}

The ABL formula for the probabilities of the results of the intermediate measurements
yields, for each
particle,
\begin{eqnarray} 
\label{CFS}
\nonumber
{\rm \bf P}_A = 1 ,\\
{\rm \bf P}_B = 1 ,\\
\nonumber{\rm \bf P}_A + {\rm \bf P}_B + {\rm \bf P}_C = 1 .
\end{eqnarray}
Or, using my definition, for each particle there are three {\it
  elements of reality}: the particle is inside box $A$,  the particle
is inside  box
$B$,  the particle is inside boxes $A$, $B$ and $C$.

A theorem in the TSQT  (Ref. \cite{AV91},  p. 2325)  says
that a  weak measurement,  in a situation in which the result of a
usual (strong)
measurement is known with certainty,   yields the same result. Thus,
from (\ref{CFS}) it follows:
\begin{eqnarray} 
\label{wv}
\nonumber({\rm \bf P}_A)_w = 1 ,\\
({\rm \bf P}_B)_w = 1 ,\\
\nonumber({\rm \bf P}_A + {\rm \bf P}_B + {\rm \bf P}_C)_w = 1 .
\end{eqnarray}
Since  for any variables,  $(X+Y)_w = X_w +Y_w$  we can  deduce that
$({\rm \bf P}_C)_w = -1$.

 Similarly,  for the   ``number operators'' such as  
${\cal N}_A \equiv \Sigma_{i=1}^{i=N}{\rm \bf P}_A^{(i)}$,
where ${\rm \bf P}_A^{(i)}$ is the projection operator on the box $A$
for a particle $i$, we obtain:
\begin{eqnarray} 
\nonumber({\cal N}_A)_w = N ,\\
({\cal N}_B)_w = N ,\\
\nonumber({\cal N}_C)_w = -N ,
\end{eqnarray}

In this rare situation  the ``weak measurement'' need not be very weak: a usual measurement of pressure is a weak
measurement of the number operator. Thus, the time-symmetrized
formalism yields  surprising result (\ref{p-p}): the pressure measurement in
box $C$ is negative! Its value equals minus the pressure measured in
the  boxes $A$ and $B$.

The analysis of ``elements of reality''  in this example which are
clearly counterfactual statements (in actual world the measurements, results
of which are quoted in (\ref{CFS}), have not been performed) yields a tangible fruit: a shortcut
for calculation of the expected outcome of an actual
measurement.\footnote{This example answers the criticism of Mermin
  \cite{Me-rub} quoted by Kastner \cite{Ka-re} in the context of my
  work. According to this criticism the elements of reality I defined
  are ``rubbish -- they have nothing to do with anything''.} 
This outcome is surprising and  paradoxical. Indeed,  a usual device for
measuring an observable which has only positive eigenvalues yields a
negative value, the weak value in this rare pre- and post-selected
situation. 

There are other paradoxical aspects discussed in relation to this
example. The first paradoxical  issue which was discussed \cite{AAD} reminds
contextuality. Consider an observable $X$ which tells us
the location of the particle: is it in box $A$, $B$, or $C$. The
eigenstate of this observable corresponding to finding the particle in
$A$ is identical to the eigenstate of the projection operator on $A$:
$~|X=A\rangle~ = ~|{\rm \bf P}_A =1\rangle$. However, in this example
there is no elements of reality $X=A$ (if we measure $X$ by opening
all boxes at time $t$ we have only the probability 1/3 to find the particle
 inside box $A$) in spite of the fact that  ${\rm \bf P}_A =1$
is an element of reality. Finally, the
paradoxical aspect of the three-box example  which was analyzed by
Kastner I shall analyze in the next section.

\section{Kastner's analysis of the three-box example.}

In the three-box example there are two elements of reality for the
{\it same} particle: ``the particle is inside box $A$'', and ``the
particle is inside box $B$''. 
Kastner \cite{Ka-f} considers this situation as a paradox which she resolves by
rejecting the legitimacy of my concept of elements of reality. She
does not mention at all my resolution of the ``paradox''. Elements of
reality are counterfactual statements. To be more explicit, ``the
particle is inside box $A$'' means that if the particle is searched in
box $A$ (and if it is not searched in box $B$!) then it is certain
that the particle would be found in box $A$. Obviously, the two
elements of reality cannot be considered together. Each element of
reality assumes that antecedent of the counterfactual statement, which
is the other element of reality, is false. Thus, both elements of
reality exist separately, but we should not conclude from this that
there is an element of reality consisting of the union of the elements
of reality: the antecedent ``the particle is searched in $A$ and it
is not searched in $B$ and the particle is searched in $B$ and it is
not searched in $A$'' is logically inconsistent. The fact that we cannot
consider the union of elements of reality does not make the whole
exercise empty. We still can consider consequences of all true
elements of reality together. In particular, in the three-box example
the consequences of elements of reality (\ref{CFS}) are the statements
about weak values (\ref{wv}) and weak measurements which yield these
weak values {\em can} be performed together.

Kastner finds elements of reality ``the particle is inside box $A$'',
and ``the particle is inside box $B$''  to be ``highly peculiar and
counterintuitive''.  This is indeed so, especially because there is no
element of reality ``the particle is inside box $A$ and inside box $B$'',
as it explained above.  This peculiar situation is an example of the
failure of the ``product rule'' for pre- and post-selected elements of
reality \cite{Va-er}. From $A=a$ and $B=b$ does not follow $AB =ab$.
The element of reality ``the particle is inside box $A$ and inside box
$B$ corresponds to the definite value of the product of projection
operators: ${\rm \bf P}_A ~{\rm \bf P}_B =1$.  But in the three-box
example ${\rm \bf P}_A ~{\rm \bf P}_B =0$, in spite of the fact that
${\rm \bf P}_A =1$ and ${\rm \bf P}_B =1$.

Kastner's main objection is that the elements of reality ``the particle
is inside box $A$'', and ``the particle is inside box $B$'' cannot be
interpreted as applying to an individual system because ``being found
in box $A$ and being found in box $B$ are mutually exclusive states of
affairs''. She does not take into account that ``elements of reality''
are just counterfactual statements. She does not pay attention on the
word ``instead'' in my writings which she herself quotes in her paper:
``If in the intermediate time it was searched for in box $A$, it has to
be found there with probability one, and if, instead, it was searched
for in box $B$, it has to be found there too with probability one...''

For demonstration that Kastner's criticism is unfounded, let me repeat
here an example of a per-selected only situation \cite{Va-bs} in
which we attribute ``mutually exclusive'' properties to an individual
system.

Consider  a system of  two spin-${1\over 2}$
particles prepared, at
$t_1$, in a singlet
state 
\begin{equation}
\label{sing}
|\Psi\rangle = {1\over {\sqrt 2}}(  |{{\uparrow}}\rangle_1
|{\downarrow}\rangle_2 -   |{\downarrow}\rangle_1
|{{\uparrow}}\rangle_2).
\end{equation}
 We can predict with certainty that 
the results of measurements of spin components of the two particles
fulfill the following two relations:
\begin{equation} 
\label{pre-er1}
\{\sigma_{1x}\} + \{\sigma_{2x}\} = 0 , \\
\end{equation} 
\begin{equation} 
\label{pre-er2}
\{\sigma_{1y}\} +\{ \sigma_{2y}\} = 0 ,
\end{equation} 
where $\{\sigma_{1x}\}$ signifies the result of measurement of the
spin $x$ component of the first particle, etc.  The relations
(\ref{pre-er1},\ref{pre-er2}) cannot be tested together: the measurement of
$\sigma_{1x}$ disturbs the measurement of $\sigma_{1y}$ and the
measurement of $\sigma_{2x}$ disturbs the measurement of $\sigma_{2y}$
(not necessarily in the same way).  According to the standard approach
to quantum theory we accept that there are two matters of fact: ``the
outcomes of the spin $x$ components for the two particles have
opposite values'' and ``the outcomes of the spin $y$ components for
the two particles have opposite values'' in spite of the fact that the
statements represent ``mutually exclusive states of affairs''. If the
spin $x$ components have been measured at time $t$, we know that $y$
components of spin were not measured at time $t$. Note that if they
were measured at a later time, after the spin $x$ component
measurement, then the outcomes might not fulfill the equation
(\ref{pre-er2}). According to
Kastner's line of argumentation the application of statements
(\ref{pre-er1},\ref{pre-er2}) which I named ``generalized elements of reality''
(because they are not just about the values of observables, but about
{\em relations} between these values) to a single quantum
system should also be rejected.  However, physicists do not reject
such statements. There ane innumerable works analyzing counterfactuals
related to incompatible measurements on a single system of correlated
spin$-{1\over 2}$ particles.  Similarly, Kastner's argumentation is not valid
for the three-box example.

\section{Quantum counterfactuals}

I will try here to clarify my statements which were criticized in
Section 4 of Kastner's paper \cite{Ka-f}. 

First, the meaning of the quotation from my work ``indeterminism is
crucial for allowing non-trivial time-symmetric counterfactuals'' is
just the following. Time-symmetric counterfactuals are related to
time-symmetric background conditions, i.e. the state of the system is
fixed both before and after the time about which the counterfactual
statement is given. In a deterministic theory everything is fixed by
conditions at a single time and, therefore, no novel (non-trivial)
features can appear in the time-symmetric approach.

In order to clarify the meaning of my continuation: 
``Lewis's and other general philosophical
analyses are irrelevant for the issue of counterfactuals
in quantum theory''  
 let  me  quote
Lewis' ``system of weights or priorities'' for similarity relation
of counterfactual worlds
\cite{Le86}:
\begin{quotation}
  (1) It is of the first importance to avoid big, widespread, diverse
  violations of [physical] law.

(2) It is of the second importance to maximize the spatio-temporal
region throughout which perfect match of particular facts prevails.

(3) It is of the third importance to avoid even small, localized,
simple violations of law.

(4) It is of little or no importance to secure approximate similarity
of particular fact, even in matters that concern us greatly.

\hfill (Lewis,  p. 47)
\end{quotation}

This priorities might be helpful in the analysis of the truth value of
a widely discussed counterfactual: ``If Nixon had pressed the
nuclear war button, the world would be very different''. The purpose
of the priorities is to ``resolve the vagueness of counterfactuals''.
In physics context, however, the counterfactuals are not vague. (At
least, I hope that counterfactuals I have defined, are not vague.)  The
truth value of quantum counterfactuals can be calculated from the
equations of quantum theory. The above  priorities cannot help in deciding
the truth value of the counterfactual ``the outcomes of the spin  $y$ 
components measurement at time $t$ for the two particles have opposite
values'' in the world in which the two spin-${1\over 2}$ particles were
prepared, at $t_1 < t$, in a singlet state (\ref{sing}) and the spin $y$
 components were measured at time $t$, instead. Priorities (1) and
(3) are not relevant because violations of physical laws are not
considered. The counterfactual worlds are different from the actual
world not because of ``miracles'', i.e., violations of physical laws, but
because different measurements on the system are considered. And the
question about how it was decided which measurement to perform, is not
under discussion.  Priorities (2) and (4) are not relevant because
quantum theory fixes everything. In particular, there is perfect match
before the time of the measurement, $t$, and, in general, there cannot be arranged the perfect
match after $t$.

We do not have the freedom of interpretation in the framework of
quantum counterfactuals after {\em defining} the similarity criteria.
For the case of pre-selected counterfactuals it is simply the identity
of quantum description of the system before the measurement and this
is not controversial. For time-symmetrized counterfactuals there is no
consensus. I have my definition. Its advantage that it yields the
standard definition as a particular case for pre-selected only
situation and it allows us to analyze and derive useful results for
pre- and post-selected quantum systems. I am aware of other proposals
\cite{grif,fin}. Each proposal should be judged according to
consistency and usefulness for the purpose it has been defined.  The
success or failure of various definitions of similarity criteria of
counterfactuals in exact sciences is not measured by maximizing
priorities (1)-(4), but by its effectiveness in the framework of a
particular theory. The priorities (1)-(4) are relevant outside the
framework of exact sciences, where we have no laws which determine
unambiguously the truth values of counterfactual statements.

Contrary to Kastner's writing I never claimed that Lewis' theory is
not applicable in an indeterministic universe. On the contrary, I have
used  Lewis' framework of possible worlds for defining
counterfactuals in quantum theory. I only claimed that most parts of
Lewis' analysis is irrelevant because counterfactuals in the context
of quantum theory are of very specific form and the majority of
aspects discussed in the general philosophical literature on
counterfactuals are not present in the quantum case. To make things
even more clear I will add another quotation from Lewis' writings
\cite{Le86} with an example of argumentation for which I cannot find
any counterpart in the analysis of quantum counterfactuals:
\begin{quotation}
Jim and Jack quarreled yesterday, and Jack is still hopping mad. We
conclude that if Jim asked Jack for help today, Jack would not help
him. But wait: Jim is a prideful fellow. He never would ask for help
after such a quarrel; if Jim were to ask Jack for help today, there
would have to have been no quarrel yesterday. In that case Jack would
be his usual generous self. So if Jim asked Jack for help today, Jack
would help him after all. ...

\hfill (Lewis,  p. 33)
\end{quotation}

Kastner continues by criticizing my definition of time-symmetrized
counterfactual regarding results of a measurement performed on pre-
and post-selected quantum system:
\begin{quotation}
  If it were that a measurement of an observable $A$
 has been performed at time $t$, \hbox{$t_1 < t < t_2$}, then
the probability for $A = a_i$ would be equal to $p_i$,
provided that the results of measurements performed
on the system at times $t_1$ and $t_2$ are
 fixed.
\end{quotation}
Her criticism  \cite{Ka-bs} regarding ``problematicity'' of the fixing requirement
 is answered in another paper \cite{Va-bs}. The latter
 was also criticized by Kastner \cite{Ka-re}. She claims that fixing
 the results of measurements at $t_1$ and $t_2$ is ``{\em ad hoc}
 gerrymanddering''  which relies on accidental similarity of
 individual facts''. But these facts are the physical assumptions in 
 the  pre- and post-selected situations for analysis of which the
 above concept of time-symmetrized counterfactuals has been
 introduced. Disregarding these facts is similar to deciding that
 there have been no quarrel between Jim and Jack even so the
 counterfactual statement starts with ``Jim and Jack quarreled
 yesterday ...''. The definitions in physics have no ambiguity which might
 allow such free reading of the text.

 In the present paper Kastner criticizes the syntax of the definition,
 in particular, that it reflects ``a confusion between the
 non-counterfactual and counterfactual usage of the ABL rule''.  In
 fact, I feel very unsure about the grammatical correctness of tenses
 in my definition.  Also, I was  not able to find exact philosophical
 definition according to which one can decide if a certain statement
 is ``counterfactual''. However, it seems to me that the meaning of my
 definition is unambiguous and the name counterfactual is appropriate
 in the context of situations this definition was applied for.  For
 example, in the three-box example described above, the definition is
 applied when it is known that in the actual world the observable $A$ (e.g.
 ${\rm \bf P}_A$) has not been measured.

 Kastner suggests two possible
 ``usages'' of my definition. The  difference, apart form
 using various tenses (the difference between which is beyond my
 linguistic understanding) is that only the second one includes the
 word ``instead''. This word is essential. According to my
 understanding it is implicit in every counterfactual statement, but
 maybe it is helpful to state it explicitly, modifying the definition
 to:
\begin{quotation}
  If it were that a measurement of an observable $A$
 has been performed at time $t$, \hbox{$t_1 < t < t_2$}, instead of
 whatever took place at time $t$ in the actual world, then
the probability for $A = a_i$ would be equal to $p_i$,
provided that the results of measurements performed
on the system at times $t_1$ and $t_2$ are
 fixed.
\end{quotation}

I hope this clarifies my definition and makes its meaning unambiguous,
even so grammatically it might not be perfect. Again, Kastner's
arguments presented in her other paper \cite{Ka-bs} that this usage of
my definition is ``generally incorrect'' have been answered in detail
elsewhere \cite{Va-bs}. Here I want only to comment on Kastner's
concluding sentence in which she writes: ``[Vaidman's] definition,
as it stands, is grammatically incorrect in a way that reflects its
lack of clarity and rigor with respect to the physically crucial point
concerning which measurement has {\it actually} taken place''.
According to my definition of time-symmetrized counterfactuals the
measurement performed at time $t$ is not ``the physically crucial point'',
on the contrary, it plays no role in calculating the truth value of
the counterfactual statement; I have noted this feature of my
definition in the paper \cite{Va-f} which Kastner criticized. The
counterfactual statement is about the counterfactual world in which at
time $t$ some action was performed {\it instead} of the measurement
which was performed in the actual world. Thus, the question which
measurement has been actually performed is clearly irrelevant. The
result of the measurement in the actual world does not add any
information either, because in the framework of standard quantum theory
to which the time-symmetrized formalism is applied, the results of
measurements at $t_1$ and $t_2$ (which are fixed by definition) yield a
complete description of the system at time $t$. 

\section{What does it mean: probability of a history?}

I want to add a comment about a connection to the consistent histories
approach \cite{grif1} advocated by Kastner and presented in the
Appendix to her paper. Following Cohen \cite{Co-gr}, Kastner claims
that the counterfactual usage of the ABL rule is valid only for cases
corresponding to ``consistent'' histories. Since for my counterfactuals
the ABL rule is valid always, I find this approach to be an unnecessary
limitation which prevents to see interesting results.

In addition, I have to admit that I was never been able to understand
the meaning of a basic concept in the consistent history approach:
probability of a history. A particular history associates set of
values of observables in a sequential set of times. If the meaning of
probability is the probability for this set to be the  results of the
measurements of these observables at the appropriate times, then this is
 a well defined question in the framework of standard quantum
theory. (The corresponding formula is given in the ABL paper
\cite{ABL}.) Apparently, the meaning is something different.  Indeed, in
the example considered by Kastner, she uses the following expression:
\begin{quotation}
``What is the probability that the system
is in state $C_k$ at time $t_1$, given that
it was preselected in state $D$ and post-selected in state $F$?''
\end{quotation}
What is the meaning of  ``the system is in state $C_k$? In this example
the system (up to known unitary transformation) {\it is} in  state
$D$. This is a  standard quantum state evolving towards the future. In
the framework of the TSQT  one can also associate
with the system at time $t_1$ the backward evolving state $F$, and to
say that the system is described by the two-state vector $\langle
F|~|D \rangle$. However, from the text of Kastner's paper it is obvious
that she considers something different. She writes: ``we consider a
framework in which the system has some value $C_k$ associated with an
arbitrary observable''.  As I mentioned in Section 1, quantum
observables do not possess values. Thus, I cannot understand the
meaning of Kastner's sentence: ``...  we cannot use the ABL rule to
calculate the probability of any particular value of either $A$ or $B$
at time $t_1$...'' because ``probability of a value'' is not defined.

In this paper I have clarified the meaning of the concepts from the
time-symmetrized quantum formalism: quantum counterfactuals and
elements of reality (which are particular quantum counterfactuals). I
have answered recent criticism of these concepts in this journal by
Kastner \cite{Ka-f}. Kastner has claimed that the three-box example is
a paradox arising from an invalid counterfactual usage of the ABL
rule. I have argued here that if one adopts my definition of quantum
counterfactuals, the ABL rule is valid. Peculiarities of this example
do not represent a true paradox, but the unusual features of pre- and
post-selected elements of reality, such as the failure of the product
rule \cite{Va-er}.

Current controversy can be added to the list of examples which led
Bell to suggest  abandoning the usage of the word
``measurement'' in quantum theory \cite{Bell}.  However, I do not think that abstaining
from using problematic concepts is the most fruitful approach. I
believe that physical and philosophical concepts which are vague and
ambiguous should continue be under discussion until the concepts and
the structure of the physical theory will be clear. I hope that
current discussion brings us closer to constructing solid foundations
for quantum theory.

 This research was supported in part by grant
471/98 of the Basic Research Foundation (administered by the Israel
Academy of Sciences and Humanities).


\end{document}